\documentclass[reprint,frontmatterverbose,showpacs,showkeys,aps]{revtex4-1}

\usepackage{graphicx}
\usepackage{dcolumn}% Align table columns on decimal point
\usepackage{bm}
\usepackage{amssymb}
\usepackage{amsmath}
\usepackage{subcaption}
\usepackage{amsthm}
\usepackage{undertilde}
\usepackage{cases}

\usepackage{hyperref}
\hypersetup{
    colorlinks=true,
    linkcolor=blue,
    citecolor=blue,
    filecolor= magenta,      
    urlcolor= magenta %cyan,
}

\newtheorem{thm}{Theorem}%[section]
\newtheorem{algo}{Algorithm}
\usepackage{hyperref}
\hypersetup{
    colorlinks=true,
    linkcolor=blue,
    citecolor=blue,
    filecolor= magenta,      
    urlcolor= magenta %cyan,
}

\newcommand{\s}{\mathbf{s}}

\newcommand{\rr}{\mathbf{r}}

\newcommand{\rp}{\mathbf{r}'}

\newcommand{\Reyuls}{{I{\relax\kern-.3em}R}}

\newcommand{\R}{\mathbb{R}}

\newcommand{\ei}{\epsilon_\infty}
\newcommand{\es}{\epsilon_s}
\newcommand{\ez}{\epsilon_0}
\newcommand{\ep}{\epsilon_p}
\newcommand{\sdiv}{{\nabla\cdot}}

\newcommand{\ee}{{\mathbf e}}
\newcommand{\pp}{{\mathbf p}}
\newcommand{\dd}{{\mathbf d}}

\begin{document}

\title{A Nonlocal Poisson-Fermi Model for Ionic Solvent}

\author{Dexuan Xie}
%\email{dxie@uwm.edu}
\thanks{Corresponding author: Dexuan Xie (dxie@uwm.edu).}
\affiliation{\small \em Department of Mathematical Sciences, University of Wisconsin-Milwaukee, Milwaukee, WI, 53201-0413, USA}

\author{Jinn-Liang Liu}
\affiliation{\small \em Department of Applied Mathematics, National Hsinchu University of Education, Hsinchu 300, Taiwan
}

\author{Bob Eisenberg}
\affiliation{\small \em Department of Molecular Biophysics and Physiology, Rush University, Chicago IL 60612, USA
}

\author{L. Ridgway Scott}
\affiliation{\small \em Department of Computer Science,
                University of Chicago,
                Chicago, IL 60637, USA
}

\date{March 15, 2016}							% Activate to display a given date or no date

\begin{abstract}
We propose a nonlocal Poisson-Fermi model for ionic solvent that includes ion size effects and polarization correlations among water molecules in the calculation of electrostatic potential. It includes the previous Poisson-Fermi models as special cases,  and its solution is the convolution of a solution of the corresponding nonlocal Poisson dielectric  model with a  Yukawa-type kernel function. Moreover, the Fermi distribution is shown to be a set of optimal ionic concentration functions in the sense of minimizing an electrostatic potential free energy. Finally, numerical results are reported to show the difference between a Poisson-Fermi solution and a corresponding Poisson solution.
\end{abstract}

\pacs{41.20.Cv, 77.22.-d, 82.60.Lf,87.10.Ed }
%PACS: 
% 41.20.Cv	Electrostatics; Poisson and Laplace equations, boundary-value problems
% 87.10.Ed	Ordinary differential equations (ODE), partial differential equations (PDE), integrodifferential models
% 77.22.-d	Dielectric properties of solids and liquids (for dielectric properties of tissues and organs, see 87.19.rf)
% 82.60.Lf	Thermodynamics of solutions

\keywords{nonlocal dielectric models, Poisson-Fermi models, electrostatics, ionic concentrations, Poisson dielectric models}

\maketitle

\section{Introduction}
Ionic solutions have been studied for a very long time, usually by using the Poisson-Boltzmann equation (PBE) as a starting point. The PBE model has its successes \cite{RN6470,RN14584,RN6554,RN22,RN6569,RN57,RN10279,li2009minimization,RN25356,RN6591,RN153}, 
particularly compared to the treatment of ionic solutions by the theory of ideal (uncharged) perfect gases, found in biochemistry texts \cite{RN136,RN21556}. But
the successes of PBE are mostly qualitative because sodium and potassium ions are not the same (for example). Real ionic solutions have nonideal excess properties \cite{RN22047,Fawcett-book2004} that distinguish between types of ions (e.g., sodium and potassium) that cannot be approximated by PBE treatments in which sodium and potassium are identical point charges. The different specific properties of ions are of the greatest importance in biology and technology
 \cite{RN21806,RN23794,RN12061,Fawcett-book2004,fraenkel2014computing,RN24848,RN20989,RN13618,RN22043,RN12291,RN12325,RN22047,RN23151,morais2001energetic,RN3035,RN7120,RN25356,RN21556}
 As Nobelist Aaron Klug (over-) states the issue \cite{RN21922}: ``There is only one word that matters in biology, and that is specificity.''  Both life and electrochemical technology (e.g., batteries) depend on the difference between ions. If your nerve cells lose their ability to deal separately with sodium and potassium ions, you die, in minutes.
 
The need for more realistic theories was well known in physical chemistry nearly a century ago and the failure to make much progress has been a source of great frustration. For example, a leading monograph, in print for more than fifty years, says ``... many workers adopt a counsel of despair, confining their interest to concentrations below about 0.02 M ...'' \cite[page 302]{RN261},   an opinion restated in even more colorful language by up-to-date references as well: ``It is still a fact that over the last decades, it was easier to fly to the moon than to describe the free energy of even the simplest salt solutions beyond a concentration of 0.1M or so'' \cite[page 10]{RN12291}. These issues are discussed, and some of the innumerable references are listed in \cite{RN471, RN449,RN25486,RN21866,RN21806,RN23227,RN22902,RN23794,Fawcett-book2004,fraenkel2014computing,RN24848,RN23144,RN13618,RN22043,RN12291,RN12325,RN22047,RN23151,RN3035,RN7120,RN261,RN24070,RN682}.
The nonideal properties of ionic solutions arise because ions are not points of charge.
Rather, ions are charged molecules that also can interact with the highly polar solvent water.
Water has a complex distribution of charge, with zero net charge but significant local centers of charge.
Much of biology depends on the properties of spherical ions that have charge independent of the local electric field (sodium and potassium ions) or are customarily treated (with reasonable success) as if they have charge independent of the local electric field (chloride and calcium ions). These bio-ions --- as they might be called because of their enormous significance to biology (documented in the classical texts of physiology and biophysics \cite{RN7109,RN24022,RN13608}) --- have nonideal properties mostly because they are spheres not points and solutions made of spheres have entropy and energy quite different from solutions of points. Bio-ions have their greatest importance where they are most concentrated, in and near the electrodes of batteries and electrochemical cells, in and near ion channel proteins, ion binding proteins (including drug receptors), nucleic acids (DNA and RNA of various types), and enzymes, particularly their active sites \cite{RN22705}. Where they are most important, ion concentrations are usually more than 10 molar, often reaching 100 molar (!), more concentrated than  table salt Na$^{+}$Cl$^{-}$ (37 molar). This surprising fact is the subject 
of the reviews \cite{RN25486,RN21866,RN24848}, and is documented there for L-type calcium channels, DEKA sodium channels, and ryanodine receptor channels. Similar charge densities are found in catalytic active sites \cite{RN22705} and in Rb$^{+}$ binding sites in the KcsA potassium channel \cite{morais2001energetic}.

In the last decades, simple treatments of ionic solutions as spheres of fixed charge in a dielectric have had surprising success in describing detailed properties of complex systems, including bulk solutions \cite{RN25486,RN21866,RN24848,liu2015numerical,liu2015poisson} and ion channels, starting with approximate treatments of bulk solutions, moving to Monte Carlo simulations of all sphere systems, culminating in a variational treatment \cite{RN12061} that combines diffusion, migration, and convection using the energetic variational approach pioneered by Chun Liu \cite{liu2009introduction,RN7010} more than anyone else. 

The variational treatment, however, computes forces between all spheres and so leads to partial differential equations that are difficult to solve even numerically in three dimensions. 
Spheres cannot overfill space and the resulting saturation phenomena can be dealt with by the Fermi-like distribution derived by Jinn-Liang Liu \cite{liu2013numerical}, which was then used by Liu and Eisenberg to compute the entropy of an arbitrary mixture of spheres of any diameter in various systems \cite{liu2013correlated,RN25560,RN25592,liu2015numerical,liu2015poisson}. This Fermi-like approach describes bulk ionic solutions, calcium channels, and the gramicidin channel (in a model based on its three dimensional structure) with some success but it is based on an approximate treatment of the energy and free energy of these systems using  
Santangelo's potential model  \cite{santangelo2006computing}  that has been popularized by  \cite{PhysRevLett.106.046102} and others \cite{tresset2008generalized}.

Santangelo's model neatly encloses the near-field correlations in a far field Poisson equation and boundary conditions that allow flow when combined with a diffusion (Nernst-Planck) representation. The separation of near and far fields depends on a single screening parameter, however, and this is clearly an over-simplification, perhaps with significant limitations, particularly in the crowded situations where ions are most important. In reality, the screening includes both solvent and solute effects, neither of which can be captured by a single parameter independent of ion concentration and type. Rather, the screening effects of other ions depend on concentration, even in the (nearly) ideal case of point charged ions, and on the diameter of ions and the composition of their solution in general. In addition, the attenuation by dielectric properties of the solvent --- that might be called dielectric screening as described by the Bjerrum constant --- must be nonlocal, because the water molecules that make up the solvent are connected by a chain of hydrogen bonds. A replacement of Santangelo's model that is nonlocal is needed, and that is what we provide here.

The study of  nonlocal dielectric continuum models was started thirty years ago to account for either the polarization correlations among water molecules or the spatial-frequency dependence of a dielectric medium in the prediction of electrostatics \cite{Basilevsky2008,dielectric1985,PhysRevLett.79.3435,dai2007new,rubinstein2010effect,NonlocalTheory1,Sahin:2014aa,lrsBIBfi}.  Because of the complexity and difficulty in numerical solution,  early work was  done only on a  Lorentz nonlocal model for the water solvent  with  charges near a half-space  or a dielectric sphere containing one central  charge or multiple charges \cite{ref:nonlocontindielsuscept,rubinstein2004influence,kornyshev1978model,vorotyntsev1978model}. 
To sharply reduce the numerical complexity,
Hildebrandt {\em et al.} developed novel reformulation techniques to modify the Lorentz nonlocal model into a system of two coupled partial differential equations (PDEs) \cite{PhysRevLett.93.108104}, opening a way to solve a nonlocal model numerically 
by advanced PDE numerical techniques \cite{hildebrandt2007electrostatic,Weggler20104059}. Motivated by Hildebrandt {\em et al.}'s work,  Xie {\em et al.} adopted different reformation techniques than the ones used by Hildebrandt {\em et al.} to reformulate the Lorentz nonlocal model for water into two decoupled PDEs, and solved them  by  a fast  finite element algorithm \cite{xie2011nonlocal}. Their reformulation techniques were then applied to the construction of a new nonlocal dielectric model for protein in water   \cite{xie_volkmer2011} and a general nonlocal Poisson dielectric model for protein in ionic solvent  
\cite{xie-nonlocal-solver2012,xie-nonlocal2014}. In fact, this general nonlocal Poisson dielectric model
is the first ionic solvent model that incorporates nonlocal dielectric effects in the field of dielectric continuum modeling. It also
 provides us with  a general framework for developing various nonlocal dielectric models.
  As one of its applications,  a nonlocal modified Poisson-Boltzmann equation (NMPBE) 
  has recently been derived as part of a nonlinear 
 nonlinear and nonlocal dielectric continuum model for protein in ionic solvent  \cite{xie-nonlocal2014}.

However, none of the current ionic models incorporate both nonlocal dielectric effects and ionic size effects due to modeling and algorithmic challenges.  As the first step toward the direction of changing this situation, in this paper, we propose a nonlocal Poisson-Fermi dielectric model for ionic solution. We generalize Santangelo's model as a nonlocal Poisson-Fermi model to 
reflect both the spatial frequency of dielectric and ionic size effects in the calculation of electrostatics for an ionic solvent containing multiple ionic species. 

In particular, we show in this paper that our nonlocal Poisson-Fermi model includes Santangelo's model as a special case, and its solution is nothing but a convolution of the solution of a nonlocal Poisson dielectric model with a Yukawa-like kernel function, which is commonly used in the construction of a nonlocal dielectric model. Furthermore, we extend the Fermi-like distribution derived by Liu and Eisenberg, and prove it to be optimal in the sense of minimizing an electrostatic free energy functional.  In fact, 
one significant feature of Liu's Fermi-Poisson treatment  \cite[eq. (10)]{liu2014poisson} is to model  interstitial voids as a particle species. However, the size of each void is position-dependent, which is difficult to deal with in the derivation of a Fermi-like distribution.
 Such a difficulty has been avoided in our new Fermi-like distribution, which only requires the bulk concentrations of ions and water 
molecules and  the radii of ions and water molecules, due to our modification on the traditional electrostatic free energy functional used in the PBE  study (see \cite{lu2011poisson,0951-7715-26-10-2899} for example).
 
Our new Poisson-Fermi model turns out to be a system of $n+2$ nonlinear equations --- one fourth order elliptic partial differential equation for defining an electrostatic potential function and $n+1$ nonlinear algebraic equations for the concentration functions of  $n$ ionic species and water molecules.  Its numerical solution raises a challenge  in computational mathematics and high performance scientific computing particularly when confined spaces must be described by jump boundary conditions, as in ion channels with discontinuities of dielectric properties and coefficients. It will be studied  in our sequential papers. In this paper,  we only report numerical results
on a simple nonlocal Poisson test model, whose analytical solution was given in Xie {\em et al.}'s recent work \cite{HansXie2015b} as well as its convolution for  a dielectric ball containing multiple charges, since our purpose is  to illustrate
 one major difference between the solutions of the Poisson-Fermi and Poisson models. Here, the convolution of the solution was done with  the same Yukawa-like kernel function as the one used in our  nonlocal Poisson-Fermi model; thus, it can be regarded as a solution of a Poisson-Fermi model even  without any consideration of
 ionic size effects. The comparison tests were done by using 488 point charges coming from a protein molecule (PDB ID: 2XLZ ). The numerical results demonstrate that the Poisson-Fermi model can have a much smoother solution and a much smaller solution range than the corresponding Poisson dielectric model. Consequently,
from a sufficient condition that we obtain to guarantees a well defined Poisson-Femi model it
implies that the Poisson-Fermi model can be much more stable numerically than the Poisson dielectric model in the calculation of ion concentrations.
  
 We organize the remaining part of the paper as follows. In Section~II, we review the nonlocal Poisson dielectric model.
In Section III, we present the new nonlocal Poisson-Fermi model. In Section~IV, we derive the new nonlocal Fermi distribution. In Section~V, we present a dimensionless nonlocal Poisson-Fermi model, and
 a sufficient condition to ensure its definition. In Section~VI, we discuss one  numerical stability issue
 related in ionic concentration calculation. The conclusions are made in Section~VII.
 
\section{A nonlocal Poisson dielectric model}
We start with a short review on the derivation of a nonlocal Poisson dielectric model.
Let  $\ee$ denote an electrostatic field. When a fixed charge density function $\rho$ and a polarization charge density function $\gamma$ are given,  $\ee$ can be simply defined by Gauss's law as follows:
%\vspace{-2mm}
\begin{equation}
   \ez \sdiv  \ee(\rr) =\gamma(\rr)+\rho(\rr)  \quad \mbox{for } \rr=(x_{1},x_{2},x_{3}) \in \R^{3},
\label{eqn:beqelecstat}
%\vspace{-2mm}
\end{equation}
where $\epsilon_0$ is the permittivity of the vacuum. However, it is difficult to obtain $\gamma$ in practice.  To avoid the difficulty,  the classic linear dielectric theory \cite{ref:debye,griffiths1999introduction}
 has been established based on the linear relationships of displacement field $\dd$ and polarization field $\pp$ with $\ee$:
%\vspace{-2mm}
\begin{equation}
\label{dd-ee-pp}
\mbox{(a)} \quad \dd(\rr) = \ez \varepsilon(\rr) \ee(\rr); \quad   \mbox{(b)} \quad \pp(\rr) = \ez \chi(\rr) \ee(\rr),  
%\vspace{-2mm}
\end{equation}
where   $\dd$ and $\pp$ are defined by
\begin{equation}
\mbox{(a)} \quad    \sdiv\dd(\rr)=\rho(\rr); \quad \quad  \mbox{(b)} \quad   - \sdiv\pp(\rr)=\gamma(\rr),
\label{eqn:polarizeqn}
%\vspace{-1mm}
\end{equation}
$\varepsilon$ is the permittivity function, and $\chi$ is the susceptibility function.
Since $\ee$ is conservative, there exists an electrostatic potential function, $\Phi$, such that
\begin{equation}
\label{ee-phi}
 \ee(\rr) = - \nabla\Phi(\rr).
\end{equation}
Hence, applying the above formula and (\ref{dd-ee-pp}a) to (\ref{eqn:polarizeqn}a), we obtain the classic Poisson dielectric model:
% \vspace{-2mm}
\begin{equation}
    - \ez \sdiv ( \varepsilon(\rr) \nabla \Phi(\rr))  =\rho(\rr)
       \quad \forall \rr \in \R^{3},     
       \label{eqn:beqelecstat2}
\end{equation} 
where $\Phi(\rr)\to 0$ as $|\rr|\to \infty$, and $\Delta=\sum_{i=1}^{3}\frac{\partial^{2}}{\partial x_{i}^{2}}$ is the Laplace operator.  

It has been known that the relationship \eqref{dd-ee-pp} depends on a spatial frequency  of a dielectric medium (see \cite{ref:hididata} for example). 
To reflect this feature, the spatial frequency variable of the Fourier transform can be employed to model the spatial frequency dependence of the dielectric response, 
so that the two linear relationships of  \eqref{dd-ee-pp} can be mimicked  in the Fourier  frequency space as follows:
% \vspace{-2mm}
\begin{equation}
\label{pp-ee-dd-Fourier}
 \mbox{(a)} \quad   \widehat{\dd}(\xi) =\ez  \widehat{\varepsilon}(\xi) \widehat{\ee}(\xi);
 \quad \mbox{(b)} \quad
     \widehat{\pp}(\xi) = \ez \widehat{\chi}(\xi) \widehat{\ee}(\xi),
%  \vspace{-2mm}
\end{equation}
where $\widehat{\varepsilon}(\xi), \widehat{\chi}(\xi), \widehat{\dd}(\xi), \widehat{\pp}(\xi)$, and $\widehat{\ee}(\xi)$ denote the Fourier transforms of  $\varepsilon(\rr), \chi(\rr), \dd(\rr)$, $\pp(\rr)$, and $\ee(\rr)$,   respectively \cite{Basilevsky2008}.
Applying the inverse Fourier transform to \eqref{pp-ee-dd-Fourier}, we obtain the  nonlocal relationships of  $\dd$ and $\pp$ with $\ee$:
\begin{subequations}
\label{eqn:nononlocndrho}
\begin{eqnarray}
   \label{eqn:nononlocndrho1}
   \dd(\rr)&=& \ez \int_{\R^{3}} \varepsilon(\rr-\rr^\prime) \ee(\rr^\prime)\,d\rr^\prime,\\
    \label{eqn:nononlocndrho2}
   \pp(\rr)&=& \ez \int_{\R^{3}} \chi(\rr-\rr^\prime) \ee(\rr^\prime)\,d\rr^\prime.
\end{eqnarray}
\end{subequations}
Substituting \eqref{eqn:nononlocndrho1} and (\ref{ee-phi}) to (\ref{eqn:polarizeqn}a), we obtain the nonlocal Poisson dielectric  model:
%\vspace{-2mm}
\begin{equation}
\label{General-Poisson}
    - \ez \sdiv \int_{\R^{3}} \varepsilon(\rr-\rr^\prime) \nabla\Phi(\rr^\prime)\,d\rr^\prime   =\rho(\rr)  \quad  \forall \rr \in \R^{3},
\end{equation}
where $\Phi(\rr)\to 0$ as $|\rr|\to \infty$.
In particular, following  Debye's (temporal)  frequency dependent permittivity function (see \cite[page 100]{Basilevsky2008} and \cite{ref:hididata} for example), we set 
$ \widehat{\varepsilon}$  in the expression
\begin{equation}
\label{eqn:debyeperm}
\widehat{\varepsilon}(\xi)=\epsilon_{\infty}+\frac{\epsilon_s-\epsilon_{\infty}}{1+\lambda^{2} |\xi|^2},
\end{equation}
where $\varepsilon_{s}$ and $\varepsilon_{\infty}$ are the static and optic values corresponding  
to the spatial frequencies $|\xi|=0$ and $|\xi| \rightarrow \infty$, respectively, $\epsilon_s >  \epsilon_{\infty}$, and  $\lambda$ is a parameter for characterizing
the spatial frequency of the water solvent as a dielectric medium (or
 the polarization correlations of water molecules) \cite{HildebrandtThesis2005,YadaIntermolecular}. 
   The inverse Fourier transform of $\widehat{\varepsilon}$ gives the commonly-used nonlocal permittivity function 
\begin{equation}
\label{epsilon-water}
    \varepsilon(\rr) = \epsilon_{\infty}\delta(\rr) +(\epsilon_s-\epsilon_{\infty})Q_{\lambda}(\rr)
\quad \forall \rr\in \R^{3}, 
\end{equation}
where $\delta$ denotes the Dirac-delta distribution at the origin  \cite{Rudin:1991fk}, and $Q_{\lambda}$ is given by
\[  Q_{\lambda}(\rr)=\frac{e^{-|\rr|/\lambda}}{4\pi \lambda^2 |\rr|}.\] 

Applying \eqref{epsilon-water} to \eqref{General-Poisson}, we obtain the nonlocal Poisson dielectric model: For $ \rr \in \R^{3},$
\begin{equation}
\label{nonlocal-Poisson}
\hspace*{-2mm} - \ez \left[ \ei \Delta \Phi(\rr) +  (\es-\ei)\sdiv  (\nabla\Phi\ast Q_\lambda)(\rr)\right]=\rho(\rr),
\end{equation}
where $\Phi(\rr)\to 0$ as $|\rr|\to \infty$, and $\nabla \Phi \ast Q_{\lambda}$ denotes the convolution of  $\nabla \Phi$ with $O_{\lambda}$, which  is defined by
\[ (\nabla \Phi\ast Q_{\lambda})(\rr)= \int_{\mathbb{R}^3}Q_{\lambda}(\rr -\rp) \nabla \Phi(\rp)\mathrm{d}\rp.\]

Furthermore, by the derivative properties of the convolution  \cite[Theorem~6.30, Page 171]{Rudin:1991fk}, the nonlocal Poisson dielectric model \eqref{nonlocal-Poisson} can be reformulated in the form
\begin{equation}
\label{nonlocal-Poisson2}
 - \ez  \Delta \left[ \ei \Phi +  (\es-\ei)  (\Phi\ast Q_\lambda)\right]=\rho(\rr), \; \rr \in \R^{3}.
\end{equation}

One interesting issue on the study of the nonlocal Poisson dielectric model comes from the selection of parameter $\lambda$.
Many studies were done for different ionic solvents and different applications (see \cite{RN471,RN20989,RN13618,RN679,RN682},  \cite[Figure~2.1]{xie2011nonlocal}, and \cite{HildebrandtThesis2005,hildebrandt2007electrostatic} for example), showing that a value of $\lambda$ can vary from 3 to 25.

While a value of $\lambda$ can be selected experimentally, it  can also be estimated theoretically by a formula 
to yield a reference value for experiments. To get such a formula,
  we rewrite $Q_{\lambda}$  as   
\[   Q_{\lambda}(\rr) = \frac{1}{\lambda^{2}} H(\rr) \quad \forall \rr \in \R^{3} \quad \mbox{with } \quad
      H=\frac{e^{-|\rr|/\lambda}}{4\pi |\rr|},\]
where $H$ is the Yukawa kernel function \cite{ref:hydroelecton}, which satisfies the distribution equation
\begin{equation}
\label{H-eq-1}
     -\Delta H(\rr) + \frac{1}{\lambda^2} H(\rr) =\delta(\rr), \quad  \rr \in \mathbb{R}^{3}.
\end{equation}

We recall that a Debye-H\"uckel equation for a symmetric 1:1 ionic solvent is defined by
\begin{equation}\label{local-MDH-def}
\displaystyle-\es\Delta u(\rr)  + {\kappa}^2 u(\rr)  = \frac{10^{10}e_{c}^{2}}{\ez k_{B}T} z \delta(\rr),\quad  \rr\in \mathbb{R}^3,
\end{equation}
where $k_B$ is the Boltzmann constant,  $T$ is the absolute temperature, $e_{c}$ is the elementary charge, $z$ is 
the charge number at the origin, and $\kappa$ is given by
\begin{equation}\label{kappa-def}
  \kappa=\frac{e_{c}}{10^{8}\sqrt{5}}\left(\frac{N_{A}I_{s}}{\ez k_{B}T}\right)^{1/2}
\end{equation} 
with $N_{A}$ being the Avogadro number ($N_{A}=6.02214129\times10^{23}$) and $I_{s}$ the ionic solvent strength. 
Clearly, the Debye-H\"uckel equation is reduced to \eqref{H-eq-1} in the case that  $z={\ez\es k_{B}T}/({10^{10}e_{c}^{2}}),$ and $\lambda={\sqrt{\es}}/{\kappa}$, from which we obtain  a formula for estimating $\lambda$:
\begin{equation}\label{lambda-def}
  \lambda=\frac{10^{8}\sqrt{5}}{e_{c}}\left(\frac{\ez\es k_{B}T}{N_{A}I_{s}}\right)^{1/2}.
\end{equation} 
Here  $\lambda$ has the length unit in angstroms (\AA) since \eqref{local-MDH-def} is in the dimensionless form produced 
by using the length unit in angstroms under the SI unit system.  

\begin{figure}
%  \vspace{-5pt}
  \begin{center}
    \includegraphics[width=0.45\textwidth]{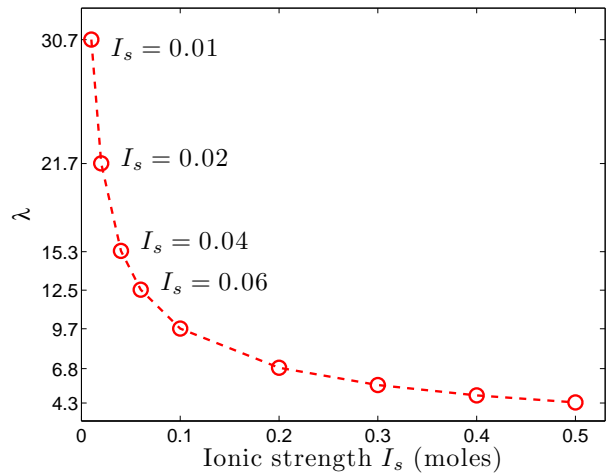}
 %  \vspace{-15pt}
  \caption{\label{lambda-Is} The correlation length parameter $\lambda$ predicted by formula \eqref{lambda-def} as a decreasing function of  $I_{s}$.
  }
    \end{center}
\end{figure}

By the formula \eqref{lambda-def} with the parameter values of $k_{B}, e_{c},\ez$ and $T$ given in \cite[Table~1]{xiePBE2013}, $\lambda$ was found to have the range $4.3 \leq \lambda \leq 30.7$ for $0.01 \leq I_{s} \leq 0.5$ as displayed in Figure~\ref{lambda-Is}. Further studies will be done on a proper selection of $\lambda$ in our sequential work. 

\section{The nonlocal Poisson-Fermi model}
In this section, we derive a nonlocal Poisson-Fermi model for computing the convolution of $\Phi$ with respect to the Yukawa-like kernel function  $Q_\lambda$. We then show that the electrostatic potential $\Phi$ can be split into two component functions ---
 one for permittivity correlations among water molecules and the other one for ionic size effects.

Doing the convolution of $\Phi$ on the both sides of \eqref{H-eq-1}, we find that $\Phi$ can be expressed in the form
\begin{equation}
\label{phi-def1}
  \Phi(\rr)= - \lambda^{2}\Delta W(\rr) +W(\rr),\quad  \; \rr \in R^{3},
\end{equation}
where  we have set $W=\Phi \ast Q_{\lambda}$, and used the derivative identity
$\Phi\ast\Delta Q_{\lambda}=  \Delta( \Phi \ast Q_{\lambda}).$ The above expression can lead to a new way for us to calculate the electrostatic potential function $\Phi$ provided that we can construct an equation for $W$ to calculate both $W$ and $\Delta W(\rr)$ quickly. 

In general, the convolution $\Phi\ast Q_\lambda$ can be 
much more smooth than $\Phi$ without  involving any singularity over the whole space $\R^{3}$. For example, 
Figure~\ref{W-phi-compare} shows that the convolution $\Phi\ast Q_{\lambda}$ has smoothed  a strongly singular $\Phi$, and sharply reduced the range of $\Phi$. Hence, an equation for $W$ should be much easier to solve numerically and should give a much more   accurate numerical solution than the equation of $\Phi$. 

We now produce an equation for $W$ from the nonlocal dielectric model \eqref{nonlocal-Poisson2}.
With \eqref{phi-def1}, we can reformulate the expression $\ei \Phi +  (\es-\ei)  (\Phi\ast Q_\lambda)$ in terms of $W$ as follows:
\begin{equation*}
\label{identity2}
 \ei \Phi +  (\es-\ei)  (\Phi\ast Q_\lambda)=\es W - \ei \lambda^{2}\Delta W.
\end{equation*}
Let $l_c^{2}=\frac{\ei}{\es}\lambda^{2}$. Applying the above expression to the nonlocal model  \eqref{nonlocal-Poisson2} yields an equation for $W$ as follows:
\begin{equation}
\label{nonlocal-Poisson-Fermi}
- \ez\es \Delta \left[W(\rr) -  l_{c}^{2} \Delta W(\rr) \right]=\rho(\rr), \quad \rr \in \R^{3},
\end{equation}
where $W\to 0$ and $\Delta W\to 0$ as $|\rr|\to \infty$, which can
be followed from    \eqref{phi-def1} and  $\Phi\to 0$ as $|\rr|\to \infty$.

As a special case, setting $\ei=\es$ reduces \eqref{nonlocal-Poisson-Fermi} to
\begin{equation}
\label{local-Poisson-Fermi}
- \ez\es \Delta \left[W(\rr) -  \lambda^{2}\Delta W(\rr) \right]=\rho(\rr), \quad \rr \in \R^{3},
\end{equation}
where $W\to 0$ as $|\rr|\to \infty$. 

Furthermore, when $\lambda=0$, the model \eqref{nonlocal-Poisson-Fermi} is reduced to
 the classic Poisson model:
\begin{equation}
\label{local-Poisson}
 -  \ez \es \Delta \Phi(\rr) =\rho(\rr) \quad  \forall \rr \in \R^{3},
\end{equation}
where  $\Phi(\rr)\to 0$ as $|\rr|\to \infty$.

From the above description it can be seen that the solutions of  \eqref{nonlocal-Poisson-Fermi} and \eqref{local-Poisson-Fermi} are  the convolutions of the solutions of nonlocal Poisson dielectric model \eqref{nonlocal-Poisson} and local Poisson dielectric model \eqref{local-Poisson}, respectively, with respect to the Yukawa-like kernel function $Q_{\lambda}$. 
For clarity, we will call  \eqref{nonlocal-Poisson-Fermi} and  \eqref{local-Poisson-Fermi}  {\em a nonlocal 
Poisson-Fermi model} and {\em a local Poisson-Fermi model}, respectively.

Note that our  local Poisson-Fermi model \eqref{local-Poisson-Fermi} is a significant generalization of Santangelo's 
 fourth-order model \cite{santangelo2006computing} since in Santangelo's model, the solution is simply treated as an electrostatic potential function, which is usually quite different from $W$.

Clearly,  with a sufficiently large domain $\Omega$, we can approximate \eqref{nonlocal-Poisson-Fermi}  
as the  boundary value problem:
\begin{equation}
\label{nonlocal-Poisson-Fermi4a}
 \left\{\begin{array}{cl}
  - \ez \es \Delta \left[W(\rr) - l_{c}^{2}\Delta W(\rr) \right] = \rho(\rr), & \rr \in \Omega,\\
     W(\s)=0, & \s\in \partial \Omega,\\
 \end{array}
\right.
\end{equation} 
where  $\partial \Omega$ denotes the  boundary of $\Omega$. 

Similar to what was done in \cite{XieLi2014}, we can show that the above boundary value problem has a unique solution,
and
there exists a continuous self-adjoint positive linear operator, $L$, such that the solution can be expressed in the operator form 
\begin{equation}
\label{W-form}
  W = L^{-1}\rho \quad \mbox{for } \rho\in L^{2}(\Omega),
\end{equation} 
where $L^{-1}$ denotes the inverse of $L$, and $L$ is defined by
\begin{eqnarray*}
(LW,v) &=&  \ez\es\Big[ l_{c}^{2} \sum_{i,j=1}^{3} \int_{\Omega} \frac{\partial^{2} W}{\partial x_{i} \partial x_{j}} \frac{\partial^{2}v}{\partial x_{i} \partial x_{j}} d\rr  \\
&& + \sum_{i=1}^{3}  \int_{\Omega}  \frac{\partial W}{\partial x_{i}} \frac{\partial v}{\partial x_{i}} d\rr
\Big] \quad \forall v \in H^{2}_{0}(\Omega),
\end{eqnarray*}
for $W\in H_{0}^{2}(\Omega)$. Here,  $H_{0}^{2}(\Omega)=\{v\in H^{2}(\Omega) \; | \; v(\s)=0 \; \forall \s\in   \partial \Omega \}$ with  
$H^{2}(\Omega)$ being a Sobolv space of functions with second order weak derivatives \cite{adams2003sobolev},
 $(u,v)= \int_{\Omega} u(\rr)v(\rr) d\rr$ is an inner product for the Hilbert space $L^{2}(\Omega)$, which is a set of functions satisfying
 $(v,v)<\infty$.

To simplify the numerical calculation, we set $$\Psi=-\Delta W$$ as a new unknown function to reformulate
  \eqref{nonlocal-Poisson-Fermi4a} as a system of two partial differential equations as follows: 
\begin{equation}
\label{nonlocal-Poisson-Fermi4b}
 \left\{\begin{array}{cl}
  - \ez \es \left[l_{c}^{2}\Delta \Psi(\rr) - \Psi(\rr)\right] = \rho(\rr), & \rr \in \Omega,\\
     \Delta W(\rr) + \Psi(\rr)  =0, & \rr \in \Omega, \\
     \Psi(\s)=0, \quad W(\s)=0, & \s\in \partial \Omega.\\
 \end{array}
\right.
\end{equation}

Using \eqref{phi-def1} and the solution $(\Psi,W)$ of  \eqref{nonlocal-Poisson-Fermi4b},
we  then  obtain the nonlocal electrostatic potential $\Phi$ by 
\begin{equation}
\label{phi-formula}
    \Phi(\rr)=W(\rr) +  \lambda^{2} \Psi(\rr),\quad  \; \rr \in \Omega.
\end{equation}

To understand the physical meaning of $\Psi$, we can use the multiplication properties of convolution to get  
\begin{eqnarray}
\label{psi-formula}
   \Psi(\rr) &=& - \Delta W(\rr) = - \Delta (\Phi \ast Q_{\lambda})(\rr) 
   = - (\Delta \Phi \ast Q_{\lambda})(\rr) \nonumber \\
   &=& \frac{1}{\ez\es} (\rho \ast Q_{\lambda})(\rr) + \frac{\es-\ei}{\ei}(W\ast Q_{\lambda})(\rr).
\end{eqnarray}
In the case of the local Poisson model \eqref{local-Poisson} (i.e., $\ei=\es$), the above expression is simplified as
\begin{eqnarray*}
 \Psi(\rr) &=& - \Delta W(\rr) = - \Delta (\Phi \ast Q_{\lambda})(\rr) \\
 &=& - (\Delta \Phi \ast Q_{\lambda})(\rr) =\frac{1}{\ez\es} (\rho \ast Q_{\lambda})(\rr).
\end{eqnarray*} 
Hence, when the charge density function $\rho$ is estimated in terms of ionic concentration functions $c_{i}$ for $i=1,2,\ldots,n$ for a solvent containing $n$ different ionic species in the expression
\begin{equation}
\label{rho-formula}
    \rho(\rr) =e_{c} \sum_{i=1}^{n} Z_{i}c_{i}(\rr), \quad \rr\in \Omega,
\end{equation}
where $Z_{i}$ is the charge number  of ionic species $i$, we can use \eqref{psi-formula} to find that 
\[ \Psi(\rr) =\frac{e_{c}}{\ez\es}  \sum_{i=1}^{n} Z_{i} (c_{i} \ast Q_{\lambda})(\rr) + \frac{\es-\ei}{\ei}(W\ast Q_{\lambda})(\rr). \]
This shows that  $\Psi$ can be used to reflect ionic size effects through properly selecting $c_{i}$. 
For this reason, the formula \eqref{phi-formula} becomes significantly important since it has split
the electrostatic potential $\Phi$ into two component functions, $\Psi$ and $W$, to reflect
both the ionic size effects and permittivity correlations among water molecules.

\section{A nonlocal Fermi distribution}

In this section, we derive a nonlocal Fermi distribution for an ionic solvent containing $n$ different ionic species,
and show that it leads to optimal ionic concentrations in the sense of minimizing an electrostatic free energy, which
we construct as a modification of the traditional one commonly used in the PBE study. 

Clearly, applying \eqref{rho-formula} to \eqref{W-form}, we can express the solution $W$ of the nonlocal Poisson-Fermi equation
\eqref{nonlocal-Poisson-Fermi4a} as a function of ionic concentrations $c_{i}$ for $i=1, 2, \ldots,n$ in the operator form
\begin{equation}\label{Wc-def}
 W= e_{c} \sum_{i=1}^{n}Z_{i} L^{-1}c_{i},
 \end{equation}
from which it implies that different ionic concentrations may result in different potential functions of $W$. Hence, it is interesting to
search for a set of  optimal ionic concentration functions to yield an optimal potential 
in the sense of an electrostatic free energy minimization. One key step to do so is to construct a proper electrostatic free energy functional in terms of ionic concentration functions.
 
To do so, we denote  $c_{n+1}$ as the concentration function of water molecules, and 
treat  the ions and water molecules as the hard-spherical balls
 with radius $a_{i}$ for $i=1,2,\ldots, n, n+1$.  Thus, the volume of each ball is given by $4\pi a_{i}^{3}/3$.   
In the ion channel study done in \cite{liu2015numerical}, the interstitial voids among these balls were considered in the calculation of ionic concentration functions. Similarly, to reflect the size effects of these voids, we define two void volume fraction functions, $\Gamma^{b}$ and $\Gamma(\rr)$, respectively, by the two size constrain conditions
\begin{equation}\label{size-constraint}
  \frac{4\pi}{3} \sum_{i=1}^{n+1} a_{i}^{3}c^{b}_{i} +  \Gamma^{b} = 1, \quad   \frac{4\pi}{3} \sum_{i=1}^{n+1} a_{i}^{3}c_{i}(\rr)+ \Gamma(\rr)= 1, %\quad \rr \in \Omega,
\end{equation}
where  $c^{b}_{i}$ denotes  the bath concentration  of the $i$th species for $i=1$ to $n+1$. We then follow 
what was done in the derivation of Boltzmann distribution from the PBE study \cite{BoLi2009a,li2009minimization,XieLi2014}, to 
select $c_{i}$ for $i=1,2,\ldots,n+1$ optimally as a solution of  the following electrostatic free energy minimization problem 
\begin{equation}\label{min-F}
\hspace*{-3mm}    \min \{ F(c )  |   c=(c_{1},c_{2},\ldots,c_{n},c_{n+1}) \mbox{ with  $c_{i} \in C(\Omega)$} \},
\end{equation}
where $F(c) =F_{es}(c)+F_{id}(c)+ F_{ex}(c)$  with $F_{es}$, $F_{id}$, and $F_{ex}$   being  the  electrostatic, ideal gas, and
excess energies, respectively, as defined by 
\begin{eqnarray*}
F_{es}(c)&=&\frac{e_{c}}{2}\sum_{i=1}^{n} \int_{\Omega} Z_{i} W(\rr) c_{i}(\rr) d\rr, \\ 
     F_{id}(c)&=&k_BT\sum_{i=1}^{n+1}\int_{\Omega} c_i(\rr) \left[\ln\left(\frac{c_{i}}{c_{i}^{b}}\right)-1 \right]d\rr,
\end{eqnarray*}
 and 
 \[ F_{ex}(c) = k_{B}T \int_{\Omega} \Gamma(\rr)\left[\ln\left(\frac{ \Gamma(\rr)}{\Gamma^{b}}\right)-1\right]d\rr.\]
 Here, $ \Gamma^{b}$ and $ \Gamma$ must be positive, and the excess energy is  induced from the size constrain conditions of \eqref{size-constraint}. 
 
 \begin{thm}
 The minimization problem \eqref{min-F} has a unique solution.  Moreover, the solution can be expressed in the Fermi distribution form
  \begin{equation}
\label{Boltzmann-Fermi2}
    c_{i}(\rr) = c_{i}^{b}e^{-\left[\frac{e_{c}Z_{i}}{k_{B}T} W(\rr) - \frac{4\pi a_{i}^{3}}{3} S^{trc}(\rr)\right]},
\end{equation}
where $ i=1,2,\ldots,n+1,$  $\Gamma$ and $\Gamma^{b}$ are defined in 
 \eqref{size-constraint}, $W$ is a solution of the nonlocal Poisson-Fermi model  \eqref{nonlocal-Poisson-Fermi4a},
 and $S^{trc}$ is defined by
 \[ S^{trc}(\rr) =\ln \left(\frac{ \Gamma(\rr)}{\Gamma^{b}}\right).\]
\end{thm}

{\em Proof. }
Clearly, by \eqref{Wc-def},  the electrostatic free energy $F_{es}$ can be reformulated as
\begin{equation*}\label{Fes-def}
 F_{es}(c)=\frac{e_{c}^{2}}{2} \sum_{i,j=1}^{n} Z_{i}Z_{j}\int_{\Omega}  L^{-1}c_{i}c_{j}d\rr.
 \end{equation*}
We then can find the first and 
second  Fr{\'e}chet partial derivatives of $F_{id}$, $F_{es}$, and $F_{ex}$ as follows:
\begin{eqnarray*}\label{pdFid}
\frac{\partial F_{id}(c)}{\partial c_{i}}&=&k_{B}T \ln\left(\frac{c_{i}}{c_{i}^{b}}\right),  \\
 \frac{\partial^{2}F_{id}(c)}{\partial c_{j}\partial c_{i}}&=&\left\{\begin{array}{ll}
\frac{k_{B}T}{c_i} & \quad  \ \ j=i,\\
0 & \quad  \ \ j\neq i,
\end{array}\right. \\
\label{pdFfes1}
     \frac{\partial F_{es}(c)}{\partial c_{i}}=e_{c}Z_{i}W, \quad & &
           \frac{\partial^{2}F_{es}(c)}{\partial c_{j}\partial c_{i}}=e^{2}_{c}Z_i Z_{j} L^{-1},\\
\label{PDfc0}
 \frac{\partial F_{ex}(c)}{\partial c_i}&=&-k_{B}T\frac{4\pi a_{i}^{3}}{3}\ln\left(\frac{ \Gamma(\rr)}{\Gamma^{b}}\right),\\
   \frac{\partial^2 F_{ex}(c)}{\partial c_j \partial c_i} &=& k_{B}T\left(\frac{4\pi}{3}\right)^{2}\frac{(a_{i}a_{j})^{3}}{\Gamma(\rr)}. 
\end{eqnarray*}
 
Combining the above partial derivatives together, we get the first  Fr{\'e}chet derivative $F^{\prime}$ in the expression 
\begin{eqnarray*}
\label{fa-derivative}
 F^{\prime}(c)v & =&\sum_{i=1}^n\int_{D_{s}}\Big[e_{c}Z_i W+k_{B}T\ln\left(\frac{c_{i}}{c_{i}^{b}}\right) \\
    & & -k_{B}T\frac{4\pi a_{i}^{3}}{3} \ln\left(\frac{ \Gamma(\rr)}{\Gamma^{b}}\right)  \Big]v_i(\rr)d\rr. 
\end{eqnarray*}

From the stationary equation $F^{\prime}(c)v =0$ it implies the system of equations: For $i=1, 2, \ldots, n+1$,
\begin{equation}\label{OE-fa}
e_{c}Z_i W+k_{B}T\ln\left(\frac{c_{i}}{c_{i}^{b}}\right)-k_{B}T\frac{4\pi a_{i}^{3}}{3} \ln\left(\frac{ \Gamma(\rr)}{\Gamma^{b}}\right)  =0.
\end{equation}

Furthermore, we can obtain that the second  Fr{\'e}chet derivative $F^{\prime\prime}$ as below: 
\begin{equation*}\label{PD} 
\begin{aligned}
F^{\prime\prime}( c)(v, v)&=e_{c}^{2}\langle L^{-1}\sum_{i=1}^n Z_iv_i, \sum_{i=1}^n Z_iv_i\rangle_{L^2(\Omega)} \\
 &+k_{B}T\sum_{i=1}^n\int_{D_s}\frac{1}{ c_i}(v_i(\rr))^2d\rr\\
& +\int_{D_{s}}\frac{k_{B}T}{\Gamma(\rr)}\left(\frac{4\pi}{3}\sum_{i=1}^n a_{i}^{3}v_i\right)^2d\rr.
\end{aligned}
\end{equation*}
From the above expression it can imply that $F^{\prime\prime}(c)$ is strictly positive. Hence, the the minimization problem \eqref{min-F} has a unique  solution. From \eqref{OE-fa} we can obtain the expressions of  \eqref{Boltzmann-Fermi2}. 
This completes the proof.

\vspace{3pt}

The term $S^{trc}$ is often referred to as {\em a steric potential}  since it describes ionic size effects caused by the ionic size constraint conditions \eqref{size-constraint} \cite{liu2015numerical}. This is the reason why the  expression of \eqref{Boltzmann-Fermi2}
can be called {\em a Fermi distribution}. 

 Specially, when all the radii $a_{i}=0$, the Fermi distribution is reduced to the Boltzmann distribution
\begin{equation*}
\label{Boltzmann}
    c_{i}(\rr) = c_{i}^{b}e^{-Z_{i}\frac{e_{c}}{k_{B}T} W(\rr)},    \quad i=1,2,\ldots,n.
\end{equation*}

In addition, setting the correlation length parameter $\lambda = 0$ (without considering any dielectric correlation effect), we obtain the classic Boltzmann distribution  
\begin{equation*}
\label{Boltzmann-classic}
    c_{i}(\rr) = c_{i}^{b}e^{-Z_{i}\frac{e_{c}}{k_{B}T} \Phi(\rr)},    \quad i=1,2,\ldots,n,
\end{equation*}
where $\Phi$ is the solution of the local Poisson dielectric equation \eqref{local-Poisson}. 

\section{A dimensionless nonlocal Poisson-Fermi model}
A combination of \eqref{OE-fa} and \eqref{rho-formula} with \eqref{nonlocal-Poisson-Fermi4a} immediately results in a system of $n+2$ equations for solving the electrostatic potential $W$ and concentration functions $\{c_{i}\}$ as follows: 
\begin{equation}
\label{nonlocal-Poisson-Fermi4c}
\hspace*{-2mm} \left\{\begin{array}{cl}
  - \ez \es \Delta \left[W - l_{c}^{2}\Delta W \right] = e_{c} \sum\limits_{i=1}^{n} Z_{i}c_{i}(\rr), \; \rr \in \Omega,&\\
  Z_{i}\frac{e_{c}}{k_{B}T} W+\ln\left(\frac{c_{i}}{c_{i}^{b}}\right)-\frac{4\pi}{3} a_{i}^{3} \ln\left(\frac{ \Gamma(\rr)}{\Gamma^{b}}\right)  =0, &  \\
 \rr \in \Omega \quad \mbox{for }  i=1, 2, \ldots, n+1, &\\
     W(\s)=0,\quad  \Delta W(\s)=0, \quad \s\in \partial \Omega. &\\
 \end{array}
\right.
\end{equation} 

In biomolecular simulation, length is measured in angstroms (\AA), and $c_{i}$ is in moles per liter. Thus, we need to convert $c_{i}$ to the number concentration (i.e., the number of ions per cubic angstroms) by
\begin{equation*}
\label{cci}
   {c}_{i} \mbox{ moles per liter} = c_{i}N_{A}10^{-27} /  \mbox{\AA}^{3}.
\end{equation*}
We then reformulate both $\Gamma(\rr)$ and $\Gamma^{b}$  as follows:
\begin{subequations}
\label{Gamma-def3}
\begin{eqnarray}\label{Gamma-def3a}
&\Gamma(\rr)=1- \frac{4\pi N_{A}}{3\times10^{27}} \sum\limits_{i=1}^{n+1} a_{i}^{3}c_{i}(\rr),& \\
 &   \Gamma^{b}=1- \frac{4\pi N_{A}}{3\times10^{27}} \sum\limits_{i=1}^{n+1} a_{i}^{3}c^{b}_{i}.&
\end{eqnarray}
\end{subequations}
Furthermore, by the variable changes
\begin{equation*}
\label{u-Phi}
  u=  \frac{e_{c}}{k_{B}T}W, \qquad \mbox{and} \quad \Psi=-\Delta u,
\end{equation*}
the nonlocal Poisson-Fermi model \eqref{nonlocal-Poisson-Fermi4c} is reformulated into the dimensionless form 
\begin{equation}
\label{nonlocal-Poisson-Fermi4d}
\hspace*{-3mm} \left\{\begin{array}{cl}
   - \es l_{c}^{2} \Delta \Psi +\es \Psi =  \frac{e_{c}^{2}N_{A}}{\ez k_{B}T10^{27}} \sum\limits_{i=1}^{n} Z_{i}c_{i} \quad \mbox{in } \Omega, &\\
   \Delta u + \Psi =0  \quad \mbox{ in } \Omega,\\
\hspace*{-13mm}  Z_{i} u+\ln(c_{i}) -\frac{4\pi}{3} a_{i}^{3} \ln \Gamma(\rr) = \ln(c_{i}^{b})  & \\
 \hspace{0.8cm}  - \frac{4\pi}{3} a_{i}^{3} \ln(\Gamma^{b}) 
    \mbox{ in } \Omega  \mbox{ for } i=1, 2, \ldots, n+1,& \\    
   u(\s)=0,\quad   \Psi(\s)=0, \quad  \s\in \partial \Omega,&\\
\end{array}
\right.
\end{equation} 
where  $\Gamma(\rr)$ and $\Gamma^{b}$ are given in \eqref{Gamma-def3}. 

To ensure  the definition of  the above system,  both 
$\Gamma(\rr)$ and $\Gamma^{b}$ are required to be positive almost every where.
In the following theorem, we present a sufficient condition to satisfy such a requirement. 

\begin{thm}
\label{ci-surf}
Let  $\Gamma(\rr)$ and $\Gamma^{b}$ be defined in \eqref{Gamma-def3}. If all  the concentration functions $c_{j}$ satisfy the range constraints
 \begin{equation}
\label{c-bound2}
  0< c_{j}(\rr) <  \frac{3\times10^{27}}{4\pi N_{A} \sum\limits_{i=1}^{n+1} a_{i}^{3}}, \quad j=1,2,\ldots, n+1,
\end{equation}
then both $\Gamma(\rr)$ and $\Gamma^{b}$ are positive. Moreover, if  
 $   0< c_{i}^{b}\leq M_{1} \mbox{ for } i=1,2,\ldots,n, $ and $  
       0<  c_{n+1}^{b}\leq M_{2},$
then a lower bound of $\Gamma^{b}$ is given by
 \begin{equation}
\label{c-bound3}
    \Gamma^{b} \geq  1 - \frac{4\pi N_{A}}{3\times10^{27}}\left[ M_{1}\sum_{i=1}^{n} a_{i}^{3}   + a_{n+1}^{3}M_{2}\right],
 \end{equation}
where $M_{1}$ and $M_{2}$ are two given upper bounds such that the above lower bound is positive. 
\end{thm}

{\em Proof. }
 By the inequality 
 $$  c_{i}(\rr) \leq  \max_{1\leq j\leq n+1}\max_{\rr\in \Omega}c_{j}(\rr),$$ we can  get
 \[  \Gamma(\rr) > 1- \frac{4\pi N_{A}}{3\times10^{27}} \sum_{i=1}^{n+1} a_{i}^{3}\max_{1\leq j\leq n+1}\max_{\rr\in \Omega}c_{j}(\rr)\]
 for $  \rr \in \Omega \mbox{ except a measure zero set}.$
 
Thus, $ \Gamma(\rr) >0$ provided that
 \[ 1-  \frac{4\pi N_{A}}{3\times10^{27}} \sum_{i=1}^{n+1} a_{i}^{3}\max_{1\leq j\leq n+1}\max_{\rr\in \Omega}c_{j}(\rr) >0,\]
from which it implies the sufficient condition \eqref{c-bound2}. 
The proof of \eqref{c-bound3} is trivial. This completes the proof.   
  
\section{Numerical stability problem}
However, the range constraint condition \eqref{c-bound2} may cause a numerical stability problem in the calculation of concentration functions. To illustrate this issue, we first construct two numerical algorithms, called Algorithms~1 and 2.  We then use them to show that the Poisson-Fermi approach is very likely to improve 
the numerical stability of the classic Poisson dielectric approach in the calculation of ionic concentrations. 

\begin{figure}[h]
        \centering
                \begin{subfigure}[b]{0.43\textwidth}
                \centering
                \includegraphics[width=\textwidth]{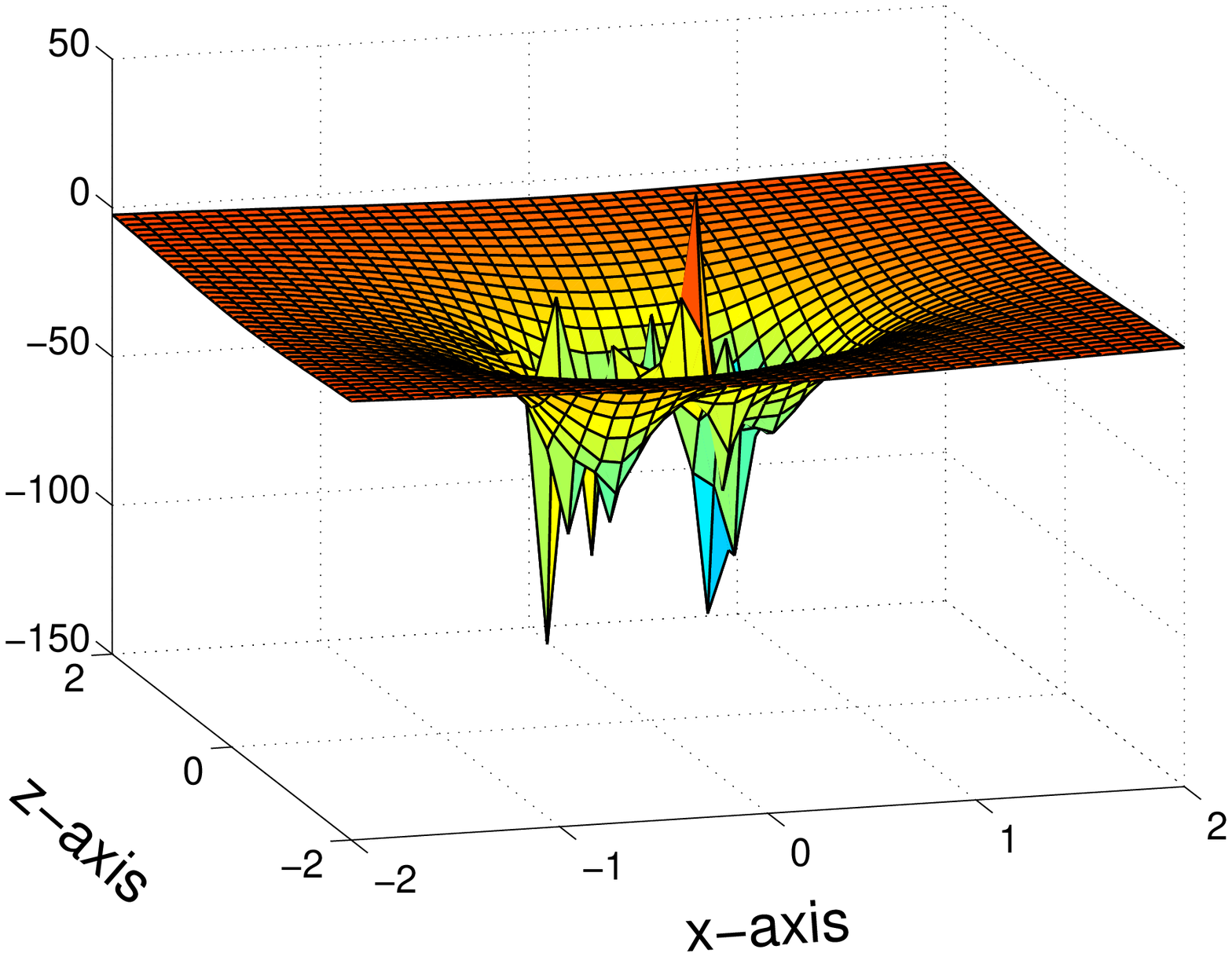}
             %   \caption{Convolution $W$ of solution $\Phi$}
                \label{4PTI-mesh1}
        \end{subfigure}
        \begin{subfigure}[b]{0.43\textwidth}
                \centering
                \includegraphics[width=\textwidth]{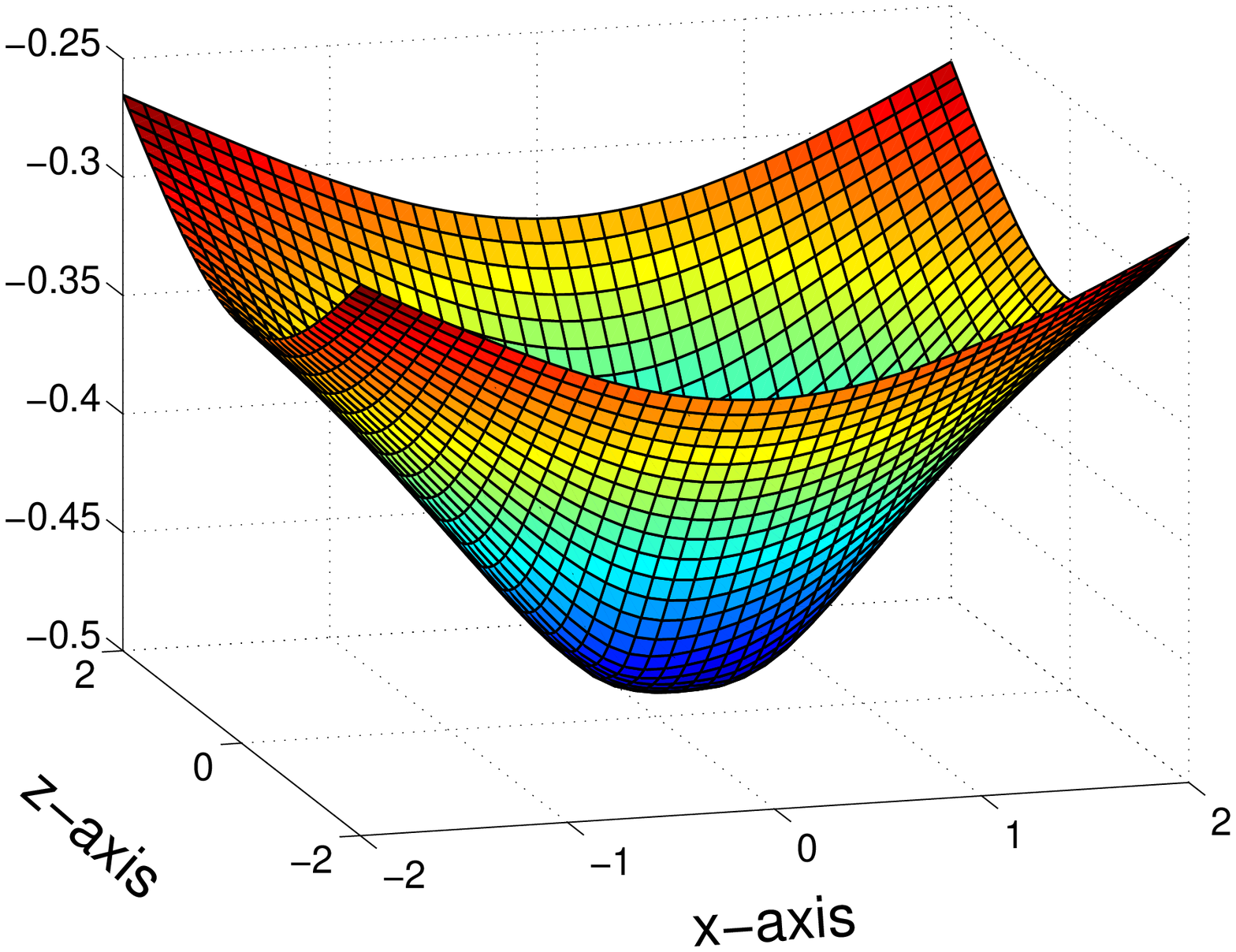}
              % \caption{Convolution $W$ of solution $\Phi$}
                \label{4PTI-mesh2}
        \end{subfigure}  
        \caption{\label{W-phi-compare}  A comparison of the solution $\Phi$ of a nonlocal Poisson dielectric test model with  the solution $W$ of  the corresponding nonlocal Poisson-Fermi model (in a view with $y=0$) for a dielectric unit  ball containing 488 point charges from a protein (PDB ID: 2LZX). 
 Here, $W=\Phi\ast Q_{\lambda}$,  and the series expressions of $\Phi$ and $W$ given in \cite{HansXie2015b}  
 were used to plot the top figure for $\Phi$ and the bottom figure for $W$.\hspace*{4.5cm}}                
\end{figure}

To construct a simple iterative scheme for computing concentration functions,
we treat the nonlinear system \eqref{nonlocal-Poisson-Fermi4d} as a differential-algebraic equation  problem. That is, 
the boundary value problem 
\begin{equation*}
\label{nonlocal-Poisson-Fermi4d-1}
 \left\{\begin{array}{cl}
   - \es l_{c}^{2} \Delta \Psi +\es \Psi =  \frac{e_{c}^{2}N_{A}}{\ez k_{B}T10^{27}} \sum\limits_{i=1}^{n} Z_{i}c_{i}, & \rr \in \Omega,\\
   \Delta u + \Psi =0, &  \rr \in \Omega,\\
   u(\s)=0,\quad   \Psi(\s)=0, & \s\in \partial \Omega,\\
\end{array}
\right.
\end{equation*} 
subject to the following  system of $n+1$ nonlinear algebraic equations: For $i=1, 2, \ldots, n+1$,
\begin{equation*}
\label{nonlocal-Poisson-Fermi4d-2}
Z_{i} u+\ln(c_{i}) -\frac{4\pi}{3} a_{i}^{3} \ln \Gamma(\rr) = \ln(c_{i}^{b}) - \frac{4\pi}{3} a_{i}^{3} \ln(\Gamma^{b}). 
  % \; \rr \in \Omega. 
\end{equation*}  
We then reformulate the above algebraic equations as
\[      c_{i}(\rr) =  \frac{c_{i}^{b}e^{-Z_{i}u(\rr)} }{(\Gamma^{b})^{{4\pi a_{i}^{3}}/{3}}}
           \left( 1- \frac{4\pi N_{A}}{3\times10^{27}} \sum_{i=1}^{n+1} a_{i}^{3}c_{i}(\rr) \right)^{\frac{4\pi a_{i}^{3}}{3}}, \] 
from which we construct a simple iterative scheme for computing concentration functions $c_{i}$ in  Algorithm~\ref{algo4ci}.

\begin{algo}
\label{algo4ci}
Let $c_{i}^{(k)}$ denote the $k$-th iterative approximation to the concentration function $c_{i}$ for $i=1,2,\ldots, n+1$.
The following four steps are carried out for $k=0, 1, 2, \ldots$ until a convergence rule is satisfied:
\begin{description}
\item[ Step 1.] Set an initial guess, $c_{i}^{(0)}$, and $k=0$.
\item[ Step 2.] Calculate the $k$-th iterate $u^{(k)}$  by solving  
the nonlocal Poisson-Fermi boundary value problem
\begin{equation*}
\label{nonlocal-Poisson-Fermi4c2}
 \left\{\begin{array}{cl}
   - \es l_{c}^{2} \Delta \Psi +\es \Psi =  \frac{e_{c}^{2}N_{A}}{\ez k_{B}T10^{27}} \sum\limits_{i=1}^{n} Z_{i}c_{i}^{(k)}, & \rr \in \Omega,\\
   \Delta u + \Psi =0, &  \rr \in \Omega,\\
   u(\s)=0,\quad   \Psi(\s)=0, \quad \s\in \partial \Omega,&\\
 \end{array}
\right.
\end{equation*} 
\item[ Step 3.] Define the update $c_{i}^{(k+1)}$ for $i=1,2,\ldots,n+1$ by the recursive formula 
\begin{eqnarray*}
\label{ci-iterate}
    && c_{i}^{(k+1)}(\rr) = \nonumber \\
    &&\hspace{0.8cm} \frac{c_{i}^{b}e^{-Z_{i}u^{(k)}(\rr)} }{(\Gamma^{b})^{{4\pi a_{i}^{3}}/{3}}}
           \left( 1- \frac{4\pi N_{A}}{3\times10^{27}} \sum_{i=1}^{n+1} a_{i}^{3}c_{i}^{(k)}(\rr) \right)^{\frac{4\pi a_{i}^{3}}{3}}.
\end{eqnarray*}  
\item[ Step 4.]   Check the convergence: \\ If $\|   c_{i}^{(k+1)} - c_{i}^{(k)}\| < \eta$ (e.g., $\eta=10^{-5}$),
stop the iteration; otherwise, increase  $k$ by one, and go back to Step~2.      
\end{description}
\end{algo}

Algorithm~2 can be constructed as a modification of  Algorithm~1 by substituting $u^{(k)}$ as $\hat{u}^{(k)}$ with 
$$\hat{u}^{(k)}=e_{c}\Phi^{(k)}/(k_{B}T),$$ 
where  $\Phi^{(k)}$ denotes a solution of   the nonlocal Poisson dielectric model \eqref{nonlocal-Poisson} using the boundary condition $\Phi=0$ on $\partial \Omega$ and the charge density $\rho(\rr)=e_{c} \sum_{i=1}^{n} Z_{i}c_{i}^{(k)}(\rr)$.  

We next use Algorithms~1 and 2 to illustrate why a Poisson-Fermi model is  more favorable than the corresponding Poisson model in the calculation  of  ionic concentrations. To do so, we only need to show that $u^{(k)}$ has a smaller range than $\hat{u}^{(k)}$ due to the fact that the value scale of the update  $c_{i}^{(k+1)}$ is mainly determined by the  factor $e^{-Z_{i}u^{(k)}} $ in Algorithm~1 or $e^{-Z_{i}\hat{u}^{(k)}}$ in Algorithms~2.
We further observe that $u^{(k)}$ is  the convolution of $\hat{u}^{(k)}$ with respect to 
the Yukawa-like kernel $Q_{\lambda}$; thus, $u^{(k)}$ can have a much smaller range than $\hat{u}^{(k)}$ since  the convolution can sharply reduce the range of a potential function in general. Therefore, 
Algorithm~1 is expected to be more stable numerically than Algorithm.

As an example, we estimated the upper bound of the constraint condition \eqref{c-bound2} for the NaCl  electrolyte.
In this numerical test, we had  $n=2$, $Z_{1}=1, Z_{2}=-1$,  and $c^{b}_{1}=c^{b}_{2}=I_{s}$. We then set
$ a_{1}=0.95,  a_{2}= 1.81,$ and $ a_{3}=1.4$ as
 the radii  of Na$^{+}$, Cl$^{-}$,  and water molecule H$_{2}$O, respectively   \cite{liu2015poisson}. In this case,
we got
  \[  0< c_{i}(\rr) <   41.6, \quad   \rr \in \Omega \quad \mbox{for } i=1,2,3, \]
confirming that the range of each ionic concentration that ensures 
$\Gamma(\rr)>0$  can be  small.    
Using \eqref{c-bound3}, we also got 
  \[   \Gamma^{b}  > 0.56 \quad \mbox{ when   $I_{s} < 2.5$ and $c_{3}^{b}<55$.} \]
 
 We next calculated the solution $\Phi$ of  a simple nonlocal Poisson test model (called Model 1 in  Xie {\em et al}'s recent work  \cite{HansXie2015b}) and its convolution $W=\Phi\ast Q_\lambda$ as an example for demonstrating
 that the convolution can sharply reduce the range of $\Phi$. 
 
 In this test, we set the solvent region $\Omega=(-2,-2)^{3}\setminus D_{p}$ with $D_{p}=\{\rr \; | \; |\rr|<1\}$, $\ep=2$, $\es=80$, $\ei=1.8$,  and $\lambda=10$.
 A set of 488 atomic charges  from a  protein with 488 atoms (PDB ID: 2LXZ) 
 were scaled to the unit ball $D_{p}$ such that each charge position has a length less than 0.8.
 The series expressions of $\Phi$ and $W$ were given in  \cite{HansXie2015b}.
 Using them, we calculated $\Phi$ and $W$  approximately as a partial sum of the series with 20 terms, which was found to have a relative error  $O(10^{-5})$ with respect to the partial sum calculated by 100 terms. Here we used the 
physical parameter values of $\epsilon_0$, $e_{c}$, $T$, and  $k_{B}$ given in  \cite[Table~1]{xiePBE2013}. 
  
Figure~\ref{W-phi-compare} shows that the Poisson-Fermi solution  $W$ is  much smoother than the Poisson solution $\Phi$, implying that  the Poisson-Fermi model can be much easier to solve numerically to high accuracy  
than the corresponding Poisson dielectric model. 

The ranges of $W$  and $\Phi$ within the solvent region $\Omega$  were found to be $[-0.4363,-0.2273]$ and $[-37.7190, -1.2484]$, respectively, confirming that the convolution has a much smaller range than  the corresponding potential function. Consequently,
 Algorithm~1 can be more numerically stable than Algorithm~2  under 
 the constrain condition \eqref{c-bound2}.
 
 \vspace{-3mm}

\section{Conclusions}

 \vspace{-3mm}

Ions always interact in water solutions, because ions are charged and so is water (although the net charge of a water molecule is zero). The interactions of ions and water and  the interactions of ions with each other have been studied extensively, first treating ions as points. Recently, the finite size of ions has been dealt with successfully in models that are easy to compute, both in flow and in mixtures, with a Fermi distribution coupled to a Poisson equation using the  Santangelo equation to link electric field near and far from ions. The Fermi distribution describes the main difference between points and finite size ions. Finite size ions cannot overfill space. Points can fill space to any density including `infinity'. 

In this work, we generalize the Santangelo equation using Xie {\em et al}'s nonlocal Poisson dielectric theory, and find a generalization that is a convolution of previous results with a Yukawa like potential. 
Our new formula for estimating the nonlocal parameter  $\lambda$ depends on and varies with ionic strength of various types of ions (mixtures in solvent). This is very different from Santagelo's model in which the parameter is a correlation length that is not specifically related to ionic strength and hence does not change with varying bulk concentrations of all ionic species. Our nonlocal Fermi distribution is new due to the specific ionic radius associated with the steric potential.
It has been shown to be optimal in the sense of minimizing an electrostatic free energy, which we construct using the bulk concentrations  and ionic size constraint conditions.
We combine our general Poisson-Fermi model with this new Fermi distribution to create a nonlocal Poisson Fermi theory
for computing both the convolution of electrostatic potential and ionic concentrations. Furthermore,  ionic concentrations are found to have limited ranges to ensure the definition of a Poisson-Fermi model. 
The convolution smooths and so results are even easier to compute than in the local theory. It also reduces the range of
a corresponding potential function sharply. Hence,
a Poisson-Fermi model is particularly valuable in the development of effective numerical algorithms for computing ionic concentrations. 

Later work will examine how well the nonlocal Poisson Fermi model fits experimental data. Moreover, this new model will be adapted to the study of biomolecules (e.g., nucleic acids and proteins) and biological applications that involve ionic flows and  concentrated ionic mixtures. Indeed, the flows, ionic mixtures, or (average global) concentrations  occur in all biological and almost all technological applications. Ions tend to be concentrated where they 
are important and Liu and Eisenberg's Poisson-Nernst-Planck-Fermi (PNPF) model has done surprisingly well in dealing with them. As an application of
our new nonlocal Fermi theory, we will develop a nonlocal PNPF model for ion channel in the future.

\vspace{-3mm}
\section*{Acknowledgement}
This work was partially supported by the
National Science Foundation, USA (DMS-1226259  to D. Xie and R. L. Scott) and the Ministry of Science and
Technology, Taiwan (MOST 103-2115-M-134-004-MY2 to J.-L. Liu).
All the authors especially express their gratitude to  the long term visitor program of the Mathematical Biosciences
Institute at the Ohio State University, USA for its supports to their visits  in the fall of 2015.

\bibliographystyle{siam}

\begin{thebibliography}{10}

\bibitem{adams2003sobolev}
{\sc R.~Adams and J.~Fournier}, {\em {S}obolev {S}paces}, vol.~140 of Pure and
  Applied Mathematics, Amsterdam). Elsevier/Academic Press, Amsterdam,,
  second~ed., 2003.

\bibitem{RN6470}
{\sc J.~Antosiewicz, J.~A. McCammon, and M.~K. Gilson}, {\em Prediction of
  p{H}-dependent properties of proteins}, J Mol Biol, 238 (1994), pp.~415--36.

\bibitem{RN6554}
{\sc N.~A. Baker}, {\em {P}oisson-{B}oltzmann methods for biomolecular
  electrostatics}, Methods in Enzymology Numerical Computer Methods, Part D,
  383 (2004), pp.~94--118.

\bibitem{RN14584}
{\sc N.~A. Baker}, {\em Improving implicit solvent simulations: a
  {P}oisson-centric view}, Curr Opin Struct Biol, 15 (2005), pp.~137--43.

\bibitem{RN471}
{\sc J.~Barthel, R.~Buchner, and M.~MŸnsterer}, {\em Electrolyte Data
  Collection Vol. 12, Part 2: Dielectric Properties of Water and Aqueous
  Electrolyte Solutions}, DECHEMA, Frankfurt am Main, 1995.

\bibitem{RN449}
{\sc J.~Barthel, H.~Krienke, and W.~Kunz}, {\em Physical Chemistry of
  Electrolyte Solutions: Modern Aspects}, Springer, New York, 1998.

\bibitem{ref:nonlocontindielsuscept}
{\sc M.~Basilevsky and D.~Parsons}, {\em Nonlocal continuum solvation model
  with exponential susceptibility kernels}, Journal of Chemical Physics, 108
  (1998), pp.~9107--9113.

\bibitem{Basilevsky2008}
{\sc V.~Basilevsky and G.~Chuev}, {\em Nonlocal solvation theories}, in
  Continuum Solvation Models in Chemical Physics: From Theory to Applications,
  B.~Mennucci and R.~Cammi, eds., Wiley, 2008, pp.~1--123.

\bibitem{PhysRevLett.106.046102}
{\sc M.~Z. Bazant, B.~D. Storey, and A.~A. Kornyshev}, {\em Double layer in
  ionic liquids: Overscreening versus crowding}, Phys. Rev. Lett., 106 (2011),
  p.~046102.

\bibitem{RN25486}
{\sc D.~Boda}, {\em {M}onte {C}arlo simulation of electrolyte solutions in
  biology: in and out of equilibrium}, Annual Review of Compuational Chemistry,
  10 (2014), pp.~127--164.

\bibitem{NonlocalTheory1}
{\sc P.~Bopp, A.~Kornyshev, and G.~Sutmann}, {\em Static nonlocal dielectric
  function of liquid water}, Phys. Rev. Lett., 76 (1996), pp.~1280--1283.

\bibitem{RN7109}
{\sc W.~Boron and E.~Boulpaep}, {\em Medical Physiology}, Saunders, New York,
  2008.

\bibitem{dai2007new}
{\sc J.~Dai, I.~Tsukerman, A.~Rubinstein, and S.~Sherman}, {\em New
  computational models for electrostatics of macromolecules in solvents},
  Magnetics, IEEE Transactions on, 43 (2007), pp.~1217--1220.

\bibitem{RN22}
{\sc M.~Davis and J.~McCammon}, {\em Electrostatics in biomolecular structure
  and dynamics.}, Chem. Rev., 90 (1990), p.~509Ð521.

\bibitem{ref:debye}
{\sc P.~Debye}, {\em Polar Molecules}, Dover, New York, 1945.

\bibitem{dielectric1985}
{\sc R.~R. Dogonadze, E.~K{\'a}lm{\'a}n, A.~A. Kornyshev, and J.~Ulstrup},
  eds., {\em The Chemical Physics of Solvation. {Part A}: Theory of Solvation},
  vol.~38 of Studies in Physical and Theoretical Chemistry, Elsevier Science
  Ltf, Amsterdam, October 1985.

\bibitem{RN21866}
{\sc B.~Eisenberg}, {\em Crowded charges in ion channels}, in Advances in
  Chemical Physics, S.~A. Rice, ed., John Wiley \& Sons, Inc., New York, 2011,
  pp.~77--223 also on the arXiv at http://arxiv.org/abs/1009.1786v1.

\bibitem{RN21806}
\leavevmode\vrule height 2pt depth -1.6pt width 23pt, {\em LifeÕs solutions are
  not ideal}, Posted on arXiv.org with Paper ID arXiv:1105.0184v1,  (2011).

\bibitem{RN23227}
\leavevmode\vrule height 2pt depth -1.6pt width 23pt, {\em A leading role for
  mathematics in the study of ionic solutions}, SIAM News, 45 (2012),
  pp.~11--12.

\bibitem{RN22902}
\leavevmode\vrule height 2pt depth -1.6pt width 23pt, {\em Life's solutions. a
  mathematical challenge.}, Available on arXiv as
  http://arxiv.org/abs/1207.4737,  (2012).

\bibitem{RN23794}
\leavevmode\vrule height 2pt depth -1.6pt width 23pt, {\em Interacting ions in
  biophysics: Real is not ideal.}, Biophysical Journal, 104 (2013),
  pp.~1849--1866.

\bibitem{RN12061}
{\sc B.~Eisenberg, Y.~Hyon, and C.~Liu}, {\em Energy variational analysis of
  ions in water and channels: Field theory for primitive models of complex
  ionic fluids}, Journal of Chemical Physics, 133 (2010), p.~104104.

\bibitem{Fawcett-book2004}
{\sc W.~R. Fawcett}, {\em Liquids, Solutions, and Interfaces: From Classical
  Macroscopic Descriptions to Modern Microscopic Details}, Oxford University
  Press, New York, 2004.

\bibitem{fraenkel2014computing}
{\sc D.~Fraenkel}, {\em Computing excess functions of ionic solutions: The
  smaller-ion shell model versus the primitive model. 1. activity
  coefficients}, Journal of chemical theory and computation, 11 (2015),
  pp.~178--192.

\bibitem{RN24848}
{\sc D.~Gillespie}, {\em A review of steric interactions of ions: {W}hy some
  theories succeed and others fail to account for ion size}, Microfluidics and
  Nanofluidics, 18 (2015), pp.~717--738.

\bibitem{RN6569}
{\sc M.~K. Gilson and B.~H. Honig}, {\em Energetics of charge-charge
  interactions in proteins}, Proteins, 3 (1988), pp.~32--52.

\bibitem{griffiths1999introduction}
{\sc D.~Griffiths}, {\em Introduction to Electrodynamics}, Prentice Hall, New
  Jersey, 3~ed., 1999.

\bibitem{HildebrandtThesis2005}
{\sc A.~Hildebrandt}, {\em Biomolecules in a structured solvent: A novel
  formulation of nonlocal electrostatics and its numerical solution}, PhD
  thesis, Saarlandes University, Saarbr\"{u}cken, Germany, February 2005.

\bibitem{hildebrandt2007electrostatic}
{\sc A.~Hildebrandt, R.~Blossey, S.~Rjasanow, O.~Kohlbacher, and H.~Lenhof},
  {\em Electrostatic potentials of proteins in water: A structured continuum
  approach}, Bioinformatics, 23 (2007), pp.~e99--e103.

\bibitem{PhysRevLett.93.108104}
{\sc A.~Hildebrandt, R.~Blossey, S.~Rjasanow, O.~Kohlbacher, and H.-P. Lenhof},
  {\em Novel formulation of nonlocal electrostatics}, Phys. Rev. Lett., 93
  (2004), p.~108104.

\bibitem{RN57}
{\sc B.~Honig and A.~Nichols}, {\em Classical electrostatics in biology and
  chemistry.}, Science, 268 (1995), pp.~1144--1149.

\bibitem{RN20989}
{\sc A.~L. Hovarth}, {\em Handbook of aqueous electrolyte solutions: physical
  properties, estimation, and correlation methods}, Ellis Horwood,, New York,
  1985.

\bibitem{RN23144}
{\sc P.~H\"unenberger and M.~Reif}, {\em Single-Ion Solvation. Experimental and
  Theoretical Approaches to Elusive Thermodynamic Quantities.}, Royal Society
  of Chemistry, London, 2011.

\bibitem{RN22705}
{\sc D.~Jimenez-Morales, J.~Liang, and B.~Eisenberg}, {\em Ionizable side
  chains at catalytic active sites of enzymes}, European Biophysics Journal, 41
  (2012), pp.~449--460.

\bibitem{ref:hididata}
{\sc U.~Kaatze, R.~Behrends, and R.~Pottel}, {\em Hydrogen network fluctuations
  and dielectric spectrometry of liquids}, Journal of Non-Crystalline Solids,
  305 (2002), pp.~19--28.

\bibitem{RN21922}
{\sc H.~Klug~in Pearson}, {\em Protein engineering: The fate of fingers},
  Nature, 455 (2008), pp.~160--164.

\bibitem{RN10279}
{\sc P.~Koehl and M.~Delarue}, {\em Aquasol: An efficient solver for the
  dipolar {P}oisson-{B}oltzmann--langevin equation}, J Chem Phys, 132 (2010),
  pp.~064101--16.

\bibitem{RN24022}
{\sc B.~Koeppen and B.~Stanton}, {\em Berne \& Levy Physiology, Updated
  Edition}, Elsevier, 6th edition~ed., 2009.

\bibitem{RN13618}
{\sc G.~M. Kontogeorgis and G.~K. Folas}, {\em Thermodynamic Models for
  Industrial Applications: {F}rom Classical and Advanced Mixing Rules to
  Association Theories}, John Wiley \& Sons, 2009.

\bibitem{ref:hydroelecton}
{\sc A.~Kornyshev and A.~Nitzan}, {\em Effect of overscreeming on the
  localization of hydrated electrons}, Zeitschrift f{{\"u}}r Physikalische
  Chemie, 215 (2001), pp.~701--715.

\bibitem{PhysRevLett.79.3435}
{\sc A.~Kornyshev and G.~Sutmann}, {\em Nonlocal dielectric saturation in
  liquid water}, Phys. Rev. Lett., 79 (1997), pp.~3435--3438.

\bibitem{kornyshev1978model}
{\sc A.~A. Kornyshev, A.~I. Rubinshtein, and M.~A. Vorotyntsev}, {\em Model
  nonlocal electrostatics. {I}}, Journal of Physics C: Solid State Physics, 11
  (1978), p.~3307.

\bibitem{RN22043}
{\sc C.~A. Kraus}, {\em The present status of the theory of electrolytes},
  Bull. Amer. Math. Soc., 44 (1938), pp.~361--383.

\bibitem{RN12291}
{\sc W.~Kunz}, {\em Specific Ion Effects}, World Scientific, Singapore, 2009.

\bibitem{RN12325}
{\sc W.~Kunz and R.~Neueder}, {\em An attempt at an overview}, in Specific Ion
  Effects, W.~Kunz, ed., World Scientific, Singapore, 2009, pp.~11--54.

\bibitem{RN22047}
{\sc K.~J. Laidler, J.~H. Meiser, and B.~C. Sanctuary}, {\em Physical
  Chemistry}, BrooksCole, Belmont CA, fourth~ed., 2003.

\bibitem{RN23151}
{\sc Y.~Levin}, {\em Electrostatic correlations: {F}rom plasma to biology},
  Reports on Progress in Physics, 65 (2002), p.~1577.

\bibitem{BoLi2009a}
{\sc B.~Li}, {\em Continuum electrostatics for ionic solutions with non-uniform
  ionic sizes}, Nonlinearity, 22 (2009), pp.~811--833.

\bibitem{li2009minimization}
\leavevmode\vrule height 2pt depth -1.6pt width 23pt, {\em Minimization of
  electrostatic free energy and the {P}oisson-{B}oltzmann equation for
  molecular solvation with implicit solvent}, SIAM J. Math. Anal, 40 (2009),
  pp.~2536--2566.

\bibitem{0951-7715-26-10-2899}
{\sc B.~Li, P.~Liu, Z.~Xu, and S.~Zhou}, {\em Ionic size effects: generalized
  boltzmann distributions, counterion stratification and modified debye
  length}, Nonlinearity, 26 (2013), p.~2899.

\bibitem{liu2009introduction}
{\sc C.~Liu}, {\em An introduction of elastic complex fluids: an energetic
  variational approach}, in Multi-Scale Phenomena In Complex Fluids: Modeling,
  Analysis and Numerical Simulation, T.~Y. Hou, C.~Liu, and J.-G. Liu, eds.,
  World Scientific, Singapore, 2009, pp.~286--337.

\bibitem{liu2013numerical}
{\sc J.-L. Liu}, {\em Numerical methods for the {P}oisson-{F}ermi equation in
  electrolytes}, Journal of Computational Physics, 247 (2013), pp.~88--99.

\bibitem{liu2013correlated}
{\sc J.-L. Liu and B.~Eisenberg}, {\em Correlated ions in a calcium channel
  model: A {P}oisson--{F}ermi theory}, The Journal of Physical Chemistry B, 117
  (2013), pp.~12051--12058.

\bibitem{RN25560}
\leavevmode\vrule height 2pt depth -1.6pt width 23pt, {\em Analytical models of
  calcium binding in a calcium channel}, The Journal of Chemical Physics, 141
  (2014), p.~075102.

\bibitem{RN25592}
\leavevmode\vrule height 2pt depth -1.6pt width 23pt, {\em
  {P}oisson-{N}ernst-{P}lanck-{F}ermi theory for modeling biological ion
  channels}, J Chem Phys, 141 (2014), p.~22D532 available on arXiv.org with
  Paper ID in arXiv.org with Paper ID arXiv:1506.06203.

\bibitem{liu2014poisson}
\leavevmode\vrule height 2pt depth -1.6pt width 23pt, {\em
  {Poisson-Nernst-Planck-Fermi} theory for modeling biological ion channels},
  The Journal of chemical physics, 141 (2014), p.~22D532.

\bibitem{liu2015numerical}
\leavevmode\vrule height 2pt depth -1.6pt width 23pt, {\em Numerical methods
  for a {Poisson-Nernst-Planck-Fermi} model of biological ion channels},
  Physical Review E, 92 (2015), p.~012711.

\bibitem{liu2015poisson}
\leavevmode\vrule height 2pt depth -1.6pt width 23pt, {\em {P}oisson--{F}ermi
  model of single ion activities in aqueous solutions}, Chemical Physics
  Letters, 637 (2015), pp.~1--6.

\bibitem{lu2011poisson}
{\sc B.~Lu and Y.~Zhou}, {\em {Poisson-Nernst-Planck} equations for simulating
  biomolecular diffusion-reaction processes {II}: size effects on ionic
  distributions and diffusion-reaction rates}, Biophysical Journal, 100 (2011),
  pp.~2475--2485.

\bibitem{morais2001energetic}
{\sc J.~H. Morais-Cabral, Y.~Zhou, and R.~MacKinnon}, {\em Energetic
  optimization of ion conduction rate by the k\&plus; selectivity filter},
  Nature, 414 (2001), pp.~37--42.

\bibitem{RN3035}
{\sc K.~S. Pitzer}, {\em Activity Coefficients in Electrolyte Solutions}, CRC
  Press, Boca Raton FL USA, 1991.

\bibitem{RN7120}
\leavevmode\vrule height 2pt depth -1.6pt width 23pt, {\em Thermodynamics},
  McGraw Hill, New York, 3rd~ed., 1995.

\bibitem{RN679}
{\sc R.~M. Pytkowicz}, {\em Activity Coefficients in Electrolyte Solutions},
  vol.~1, CRC, Boca Raton FL USA, 1979.

\bibitem{RN25356}
{\sc P.~Ren, J.~Chun, D.~G. Thomas, M.~J. Schnieders, M.~Marucho, J.~Zhang, and
  N.~A. Baker}, {\em Biomolecular electrostatics and solvation: a computational
  perspective}, Quarterly Reviews of Biophysics, 45 (2012), pp.~427--491.

\bibitem{RN261}
{\sc R.~Robinson and R.~Stokes}, {\em Electrolyte Solutions}, Butterworths
  Scientific Publications, also Dover books, 2002., London, second~ed., 1959.

\bibitem{rubinstein2010effect}
{\sc A.~Rubinstein, R.~Sabirianov, W.~Mei, F.~Namavar, and A.~Khoynezhad}, {\em
  Effect of the ordered interfacial water layer in protein complex formation: A
  nonlocal electrostatic approach}, Physical Review E, 82 (2010), p.~021915.

\bibitem{rubinstein2004influence}
{\sc A.~Rubinstein and S.~Sherman}, {\em Influence of the solvent structure on
  the electrostatic interactions in proteins}, Biophysical Journal, 87 (2004),
  pp.~1544--1557.

\bibitem{RN13608}
{\sc T.~C. Ruch and H.~D. Patton}, {\em Physiology and Biophysics, Volume 1:
  The Brain and Neural Function}, vol.~1, W.B. Saunders Company, Philadelphia,
  1973.

\bibitem{Rudin:1991fk}
{\sc W.~Rudin}, {\em Functional Analysis}, McGraw-Hill, New York, 2nd~ed.,
  1991.

\bibitem{RN7010}
{\sc R.~Ryham, C.~Liu, and L.~Zikatanov}, {\em Mathematical models for the
  deformation of electrolyte droplets}, Discrete and Continuous Dynamical
  Systems-Series B, 8 (2007), pp.~649--661.

\bibitem{Sahin:2014aa}
{\sc B.~Sahin and B.~Ralf}, {\em Nonlocal and nonlinear electrostatics of a
  dipolar coulomb fluid}, Journal of Physics: Condensed Matter, 26 (2014),
  p.~285101.

\bibitem{santangelo2006computing}
{\sc C.~D. Santangelo}, {\em Computing counterion densities at intermediate
  coupling}, Physical Review E, 73 (2006), p.~041512.

\bibitem{lrsBIBfi}
{\sc L.~Scott, M.~Boland, K.~Rogale, and A.~Fern{\'a}ndez}, {\em Continuum
  equations for dielectric response to macro-molecular assemblies at the nano
  scale}, Journal of Physics A: Math. Gen., 37 (2004), pp.~9791--9803.

\bibitem{RN136}
{\sc L.~Stryer}, {\em Biochemistry}, W.H. Freeman, New York, fourth~ed., 1995.

\bibitem{RN6591}
{\sc S.~Subramaniam}, {\em Treatment of electrostatic effects in proteins:
  Multigrid-based {N}ewton iterative method for solution of the full nonlinear
  {P}oisson-{B}oltzmann equation}, Prot. Struct. Func. Gen., 18 (1994),
  pp.~231--245.

\bibitem{RN153}
{\sc R.~Tan, T.~Truong, and J.~McCammon}, {\em Acetylcholinesterase:
  Electrostatic steering increases the rate of ligand binding}, Biochemistry,
  32 (1993), pp.~401--403.

\bibitem{tresset2008generalized}
{\sc G.~Tresset}, {\em Generalized {Poisson-Fermi} formalism for investigating
  size correlation effects with multiple ions}, Physical Review E, 78 (2008),
  p.~061506.

\bibitem{RN24070}
{\sc V.~Vlachy}, {\em Ionic effects beyond {P}oisson-{B}oltzmann theory},
  Annual Review of Physical Chemistry, 50 (1999), pp.~145--165.

\bibitem{RN21556}
{\sc D.~Voet and J.~Voet}, {\em Biochemistry}, John Wiley, Hoboken, NJ USA,
  third~ed., 2004.

\bibitem{vorotyntsev1978model}
{\sc M.~Vorotyntsev}, {\em Model nonlocal electrostatics. {II}. {S}pherical
  interface}, Journal of Physics C: Solid State Physics, 11 (1978), p.~3323.

\bibitem{Weggler20104059}
{\sc S.~Weggler, V.~Rutka, and A.~Hildebrandt}, {\em A new numerical method for
  nonlocal electrostatics in biomolecular simulations}, J. Comput. Phys., 229
  (2010), pp.~4059 -- 4074.

\bibitem{xiePBE2013}
{\sc D.~Xie}, {\em New solution decomposition and minimization schemes for
  {P}oisson-{B}oltzmann equation in calculation of biomolecular
  electrostatics}, J. Comput. Phys., 275 (2014), pp.~294--309.

\bibitem{xie-nonlocal2014}
{\sc D.~Xie and Y.~Jiang}, {\em A nonlocal modified {P}oisson-{B}oltzmann
  equation and finite element solver for computing electrostatics of
  biomolecules}.
\newblock Submitted, 2015.

\bibitem{xie2011nonlocal}
{\sc D.~Xie, Y.~Jiang, P.~Brune, and L.~Scott}, {\em A fast solver for a
  nonlocal dielectric continuum model}, {SIAM} J. Sci. Comput., 34 (2012),
  pp.~B107--B126.

\bibitem{xie-nonlocal-solver2012}
{\sc D.~Xie, Y.~Jiang, and L.~Scott}, {\em Efficient algorithms for a nonlocal
  dielectric model for protein in ionic solvent}, {SIAM} J. Sci. Comput., 38
  (2013), pp.~B1267--1284.

\bibitem{XieLi2014}
{\sc D.~Xie and J.~Li}, {\em A new analysis of electrostatic free energy
  minimization and {P}oisson-{B}oltzmann equation for protein in ionic
  solvent}, Nonlinear Analysis: Real World Applications, 21 (2015),
  pp.~185--196.

\bibitem{xie_volkmer2011}
{\sc D.~Xie and H.~Volkmer}, {\em A modified nonlocal continuum electrostatic
  model for protein in water and its analytical solutions for ionic {B}orn
  models}, Commun. Comput. Phys., 13 (2013), pp.~174--194.

\bibitem{HansXie2015b}
{\sc D.~Xie, H.~W. Volkmer, and J.~Ying}, {\em Analytical solutions of nonlocal
  {P}oisson dielectric models with multiple point charges inside a dielectric
  sphere}, Physical Review E,  (2016).
\newblock Accepted.

\bibitem{YadaIntermolecular}
{\sc H.~Yada, M.~Nagai, and K.~Tanaka}, {\em The intermolecular stretching
  vibration mode in water isotopes investigated with broadband terahertz
  time-domain spectroscopy}, Chemical Physics Letters, 473 (2009),
  pp.~279--283.

\bibitem{RN682}
{\sc J.~Zemaitis, Joseph~F., D.~M. Clark, M.~Rafal, and N.~C. Scrivner}, {\em
  Handbook of Aqueous Electrolyte Thermodynamics}, Design Institute for
  Physical Property Data, American Institute of Chemical Engineers, New York,
  1986.

\end{thebibliography}

\end{document}